\documentclass[useAMS,usenatbib]{mn2e}
\def\aap{AA}

\def\apjl{ApJL}
\def\na{NewA}

\def\mnras{MNRAS}
\def\apj{ApJ}
\def\apjs{ApJS}

\def\prd{PhysRevD}
\def\jcap{JCAP}

\def\dgemail{gilmanda@ucla.edu}

\usepackage[titletoc]{appendix}
\usepackage{graphicx}
\usepackage{float}
\usepackage{amssymb}
\usepackage{amsfonts}
\usepackage{amsmath} 
\usepackage{color}
\usepackage{wrapfig}
\usepackage{hyperref}
\usepackage{algorithm}
\usepackage{algpseudocode}
\usepackage{physics}

\usepackage{natbib}

\def\msun{{M_{\odot}}}

\def\dlos{{\delta_{\rm{los}}}}

\def\sevenonesigma{{$c = 15_{-8}^{+9}$}}
\def\eightonesigma{{$c = 12_{-5}^{+6}$}}
\def\nineonesigma{{$c = 10_{-4}^{+7}$}}

\def\seventwosigma{{$c = 15_{-11}^{+18}$}}
\def\eighttwosigma{{$c = 12_{-9}^{+15}$}}
\def\ninetwosigma{{$c = 10_{-7}^{+14}$}}

\title[strong lensing constraints on halo concentrations]{Constraints on the mass-concentration relation of cold dark matter halos with 11 strong gravitational lenses}
\author[Gilman et al.]{\parbox{\textwidth}{
		Daniel Gilman$^{1}$\thanks{\dgemail}, 
		Xiaolong Du$^{2}$,
		Andrew Benson$^{2}$,
		Simon Birrer$^{1}$,
		Anna Nierenberg$^{3}$,
		Tommaso Treu$^{1}$\\
	}
	\\
	\parbox{\textwidth}{
		$^{1}$Department of Physics and Astronomy, University of California,
		Los Angeles, CA 90095, USA\\
		$^{2}$Carnegie Observatories, 813 Santa Barbara Street, Pasadena, CA 91101, USA\\
		$^{3}$Jet Propulsion Laboratory, California Institute of Technology, 4800 Oak Grove Dr, Pasadena, CA 91109, USA
	}
}

\hypersetup{draft}
\begin{document}
	
	\voffset-.6in
	
	\date{Accepted . Received }
	
	\pagerange{\pageref{firstpage}--\pageref{lastpage}} 
	
	\maketitle	
	\label{firstpage}
	\begin{abstract}
		The mass-concentration relation of dark matter halos reflects the assembly history of objects in hierarchical structure formation scenarios, and depends on fundamental quantities in cosmology such as the slope of the primordial matter power-spectrum. This relation is unconstrained by observations on sub-galactic scales. We derive the first measurement of the mass-concentration relation using the image positions and flux ratios from eleven quadruple-image strong gravitational lenses (quads) in the mass range $10^{6} - 10^{10} \msun$, assuming cold dark matter. Our analysis framework includes both subhalos and line of sight halos, marginalizes over nuisance parameters describing the lens macromodel, accounts for finite source effects on lensing observables, and simultaneously constrains the normalization and logarithmic slope of the mass-concentration relation, and the normalization of the subhalo mass function. At $z=0$, we constrain the concentration of $10^{8} M_{\odot}$ halos $c=12_{-5}^{+6}$ at $68 \%$ CI, and $c=12_{-9}^{+15}$  at $95 \%$ CI. For a $10^{7} \msun$ halo, we obtain $68 \%$ ($95 \%$) constraints $c=15_{-8}^{+9}$  ($c=15_{-11}^{+18}$), while for $10^{9} M_{\odot}$ halos $c=10_{-4}^{+7}$  ($c=10_{-7}^{+14}$). These results are consistent with the theoretical predictions from mass-concentration relations in the literature, and establish strong lensing by galaxies as a powerful probe of halo concentrations on sub-galactic scales across cosmological distance. 
	\end{abstract}
	
	
	\section{Introduction}
	Dark matter structure in cold dark matter (CDM) cosmologies proceeds hierarchically. Small peaks in the density field collapse first, followed by the collapse of over-densities on larger scales and mergers between collapsed halos \citep{Navarro++97,Moore++99}. The scale-free nature of structure formation in CDM scenarios results in self-similar density profiles for individual dark matter halos, first pointed out by \citet{Navarro++96} (hereafter NFW). The concentration parameter, defined as the ratio of the virial radius of the halo to its scale radius $c \equiv \frac{r_{\rm{vir}}}{r_s}$, determines the density profile of NFW halos $\rho\left(r\right)$
	
	\begin{equation}
	\frac{\rho \left(r\right)}{\rho_{\rm{crit}}} = \frac{\delta_{\rm{vir}}\left(c\right)}{x \left(1+x\right)^2}
	\end{equation}
	where $\rho_{\rm{crit}}$ is the critical density of the universe today and $x = \frac{r}{r_s}$. The concentration parameter enters through the definition of the virial overdensity $\delta_{\rm{vir}}$
	\begin{equation}
	\delta_{\rm{vir}} \left(c\right) = \frac{200}{3} \frac{c^3}{\ln(1+c)  - \frac{c}{1+c}}
	\end{equation}
	where we define the definition of the virial radius of the halo as the boundary radius $r_{\rm{vir}} = r_{200}$ enclosing a mean density $200 \times \rho_{\rm{crit}}$. The function $c \left(M, z\right)$ relates the concentration of a halo to its mass and redshift, and is known as the mass-concentration relation.
	
	\citet{Navarro++97} argued that the anti-correlation between halo mass and concentration seen in N-body simulations reflects the collapse epoch of the halo, with the high concentrations of low-mass objects reflective of the higher background density of the universe at early times when the majority of these small over-densities collapsed. The logarithmic slope of the matter power-spectrum $P\left(k\right) \propto k^{n}$ is also understood to affect halo concentrations, with larger $n$ resulting in more centrally concentrated low-mass halos \citep{Eke++01}. These realizations provided a starting point for attempts at predicting the mass-concentration relation of cold dark matter halos \citep[e.g.][]{Bullock++01,Wechsler++02,Prada++12,vandenBosch++14,Ludlow++14,DiemerJoyce19}. As the various models for halo concentrations depend on the mass accretion history of dark matter halos and the matter-power spectrum, the mass-concentration relation encodes information regarding the process of dark matter structure formation in the universe. Halo concentrations also play a central role in determining the dark matter annihilation signals from dwarf galaxies \citep{Strigari++07}. 
	
	\begin{figure}
		\includegraphics[clip,trim=0cm 0cm 0cm
		0cm,width=0.45\textwidth,keepaspectratio]{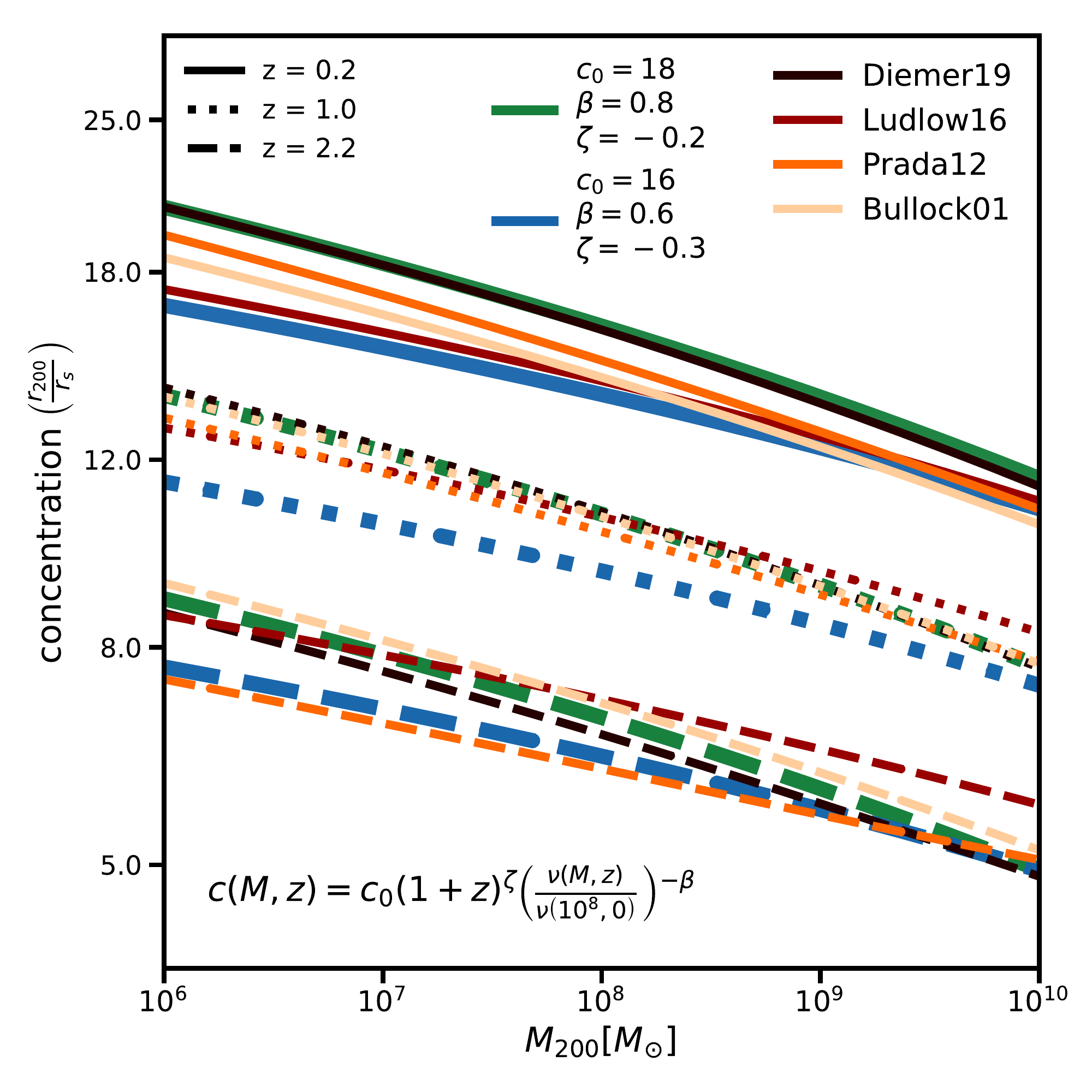}
		\caption{\label{fig:mcrelations} Mass-concentration relations from the literature as a function of halo mass and redshift, compared with the functional form for the relation in Equation \ref{eqn:mcrelation}. The parameterization of the mass-concentration relation we constrain in this work has a variable normalization $c_0$ and logarithmic slope $\beta$, with a redshift evolution modified by an empirical factor $\zeta$. Models plotted from the literature are valid in the mass range shown in the figure.}
	\end{figure}	
	
	Despite progress over the past two decades in identifying the astrophysical process shaping the mass-concentration relation, a first-principles derivation does not exist and the form of the mass-concentration relation on mass scales below $10^{9} \msun$ remains unconstrained by observations. We remedy this situation using the flux ratios and images positions from 11 quadruple-image strong gravitational lenses to constrain the mass-concentration relation on scales $\leq 10^{8} \msun$, where halos are expected to be mostly devoid of stars and gas. Strong lensing is a powerful tool for constraining the abundance and structure of dark matter halos as it measures dark matter structures directly, without relying on luminous matter to trace the dark matter \citep[e.g.][]{D+K02,Veg++14,Nierenberg++14,Birrer++17a,Hsueh++19,Gilman++19b}. In this letter, we deploy the statistical machinery developed and tested by \citet{Gilman++19} to constrain the mass-concentration relation of CDM halos on scales below $10^8 \msun$ at cosmological distance. 
	\begin{figure}
		\includegraphics[clip,trim=0cm 0cm 0cm
		0cm,width=0.45\textwidth,keepaspectratio]{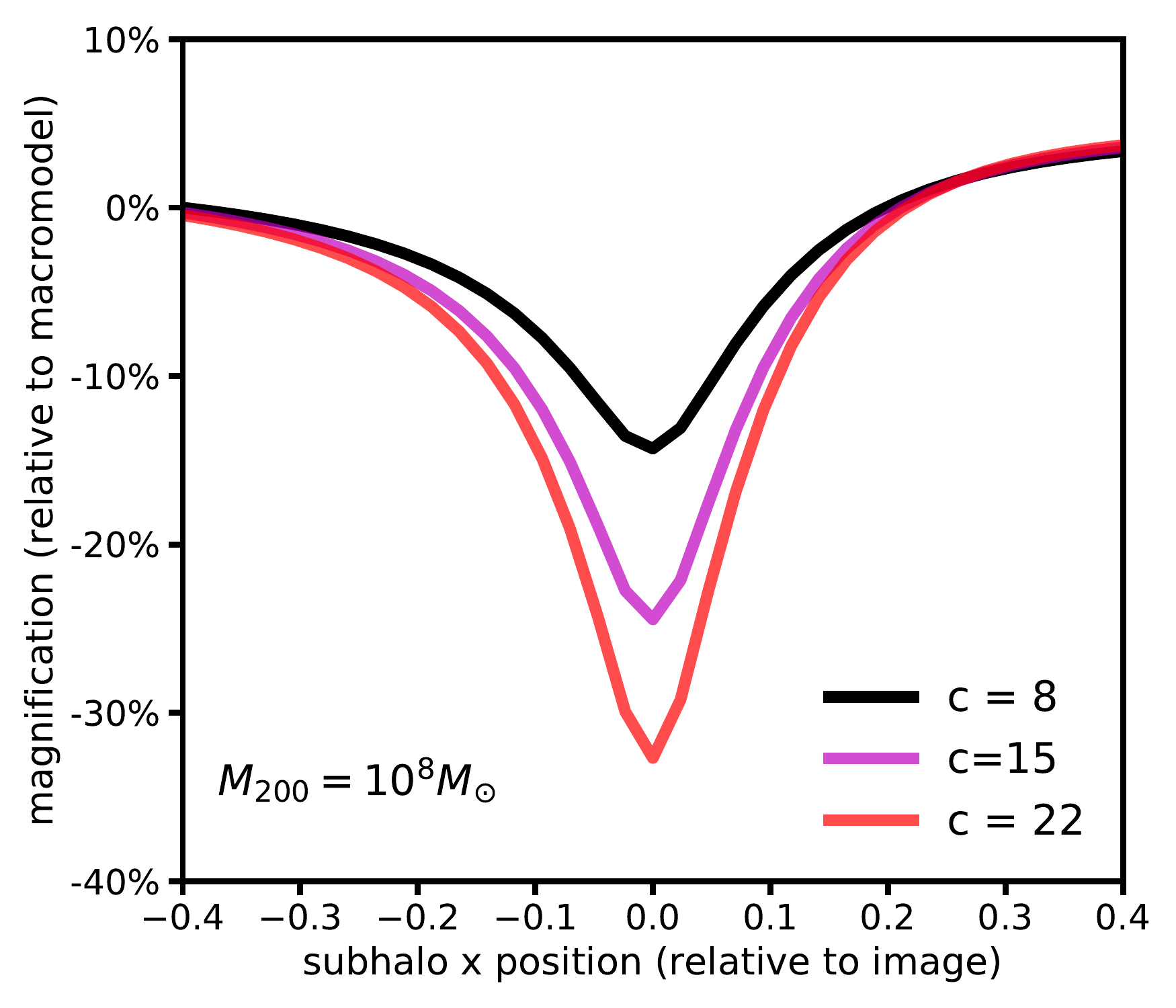}
		\caption{\label{fig:cross} Magnification perturbation cross section of a $10^{8} \msun$ halo at $z=0.5$ with various concentrations for a 30 pc background source at $z=1.5$. More concentrated halos are more efficient lenses, resulting in stronger flux perturbations.}
	\end{figure}	
	
	This letter is organized as follows. In Section \ref{sec:modeling}, we review the parameterizations of the subhalo and halo mass functions, and describe the parameterization of the mass-concentration relation we constrain in this work. We also briefly discuss the observable signatures of halo concentrations on image flux ratios, and briefly review our Bayesian inference methodology. For more detailed discussion of the mass functions and inference method, we defer to the text of \citet{Gilman++19b}. Section \ref{sec:data} discusses the data used in this analysis. In Section \ref{sec:results}, we present our main results, and we provide concluding remarks in Section \ref{sec:conclusions}. 
	
	Lensing computations are carried out using {\tt{lenstronomy}}\footnote{https://github.com/sibirrer/lenstronomy} \citep{BirrerAmara18}. Computations involving the halo mass function and the matter power spectrum are performed with {\tt{colossus}} \citep{Diemer17}. We assume a standard cosmology using the parameters from WMAP9 \citep{WMAP9cosmo} ($\Omega_m = 0.28, \sigma_8 = 0.82, h=0.7$).  
	
	\section{Modeling strategy and inference method}	
	\label{sec:modeling}
	In this section we describe the modeling of the CDM subhalo and line of sight halo mass functions, and a model for the mass-concentration relation of CDM halos expressed in terms of the peak height $\nu$. In Sections 2 and 3 of \citet{Gilman++19b} we provide additional detail regarding the Bayesian inference methodology and the mass function parameterizations. 
	
	\subsection{A model for the CDM mass-concentration relation for field halos} 
	\begin{figure}
		\includegraphics[clip,trim=0cm 0cm 0cm
		0cm,width=0.5\textwidth,keepaspectratio]{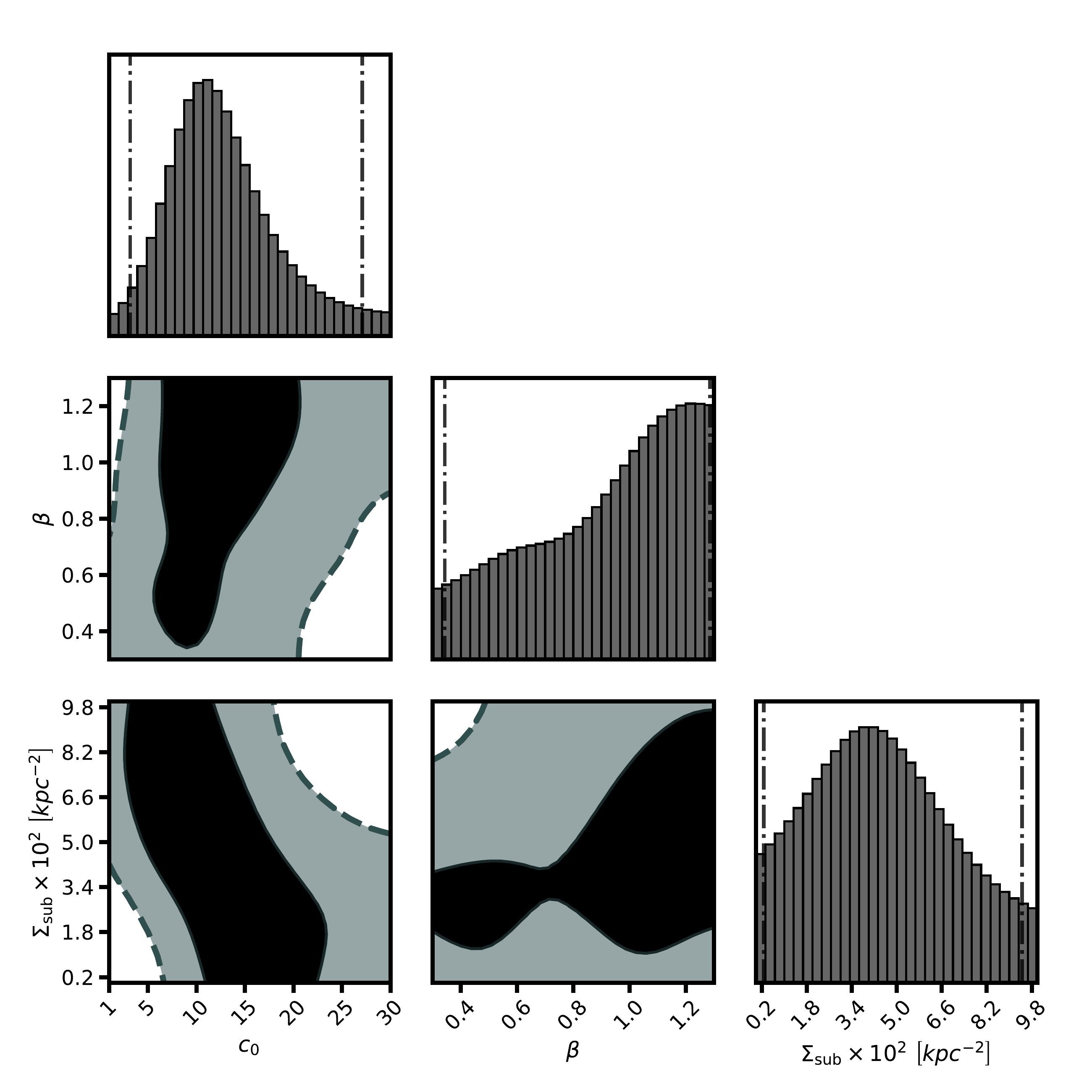}
		\caption{\label{fig:joint} Constraints on the normalization $c_0$ and the logarithmic slope $\beta$ of the mass-concentration relation in Equation \ref{eqn:mcrelation}. We include the constraints on the normalization of the subhalo mass function $\Sigma_{\rm{sub}}$, as it is covariant with both $c_0$ and $\beta$. Contours show $68\%$ and $95\%$ confidence intervals. The parameter $\zeta$, for which we use a Gaussian prior $\mathcal{N} \left(-0.25, 0.05\right)$, is unconstrained. }
	\end{figure}	
	
	The mass function for field halos\footnote{In lensing nomenclature these are also referred to as `line of sight' halos.} is parameterized in terms of the Sheth-Tormen (ST) halo mass function \citep{ST99} $\frac{d^2N}{dm  dV} \big \vert_{\rm{ShethTormen}}$
	\begin{equation}
	\label{eqn:losmfunc}
	\frac{d^2N_{\rm{los}}}{dm  dV} = \dlos \big(1+ \xi_{\rm{2halo}}\left(M_{\rm{halo}}, z\right)\big) \frac{d^2N}{dm  dV} \big \vert_{\rm{ShethTormen}}.
	\end{equation}
	The parameter $\dlos$ is an overall scaling term that accounts for a systematic shift the mean number of halos predicted by the ST mass function, and $\xi_{\rm{2halo}}$ is the two-halo term that introduces additional correlated structure around the host dark matter halo \citep{Gilman++19b}.
	
	Halos in the field by definition follow the median mass-concentration relation, which we parameterize as 
	
	\begin{equation}
	\label{eqn:mcrelation}
	c\left(M, z\right) = c_0  \left(1+z\right)^{\zeta} \left(\frac{\nu\left(M, z\right)}{\nu\left(10^8, 0\right)}\right)^{-\beta}
	\end{equation}
	with a scatter of 0.1 dex \citep{Dutton++14}. The above relation is expressed as a power-law with slope $\beta$ in terms of the peak height $\nu$ at a particular length scale $R = \left(\frac{3M}{4 \pi \rho_{\rm{m,0}} }\right)^\frac{1}{3}$
	\begin{equation}
	\nonumber\nu\left(R,z\right) = \frac{\delta_c}{\sigma \left(R,z\right)}
	\end{equation}
	where $\delta_c = 1.686$ is the threshold for spherical collapse in an Einstein de-Sitter universe, $\rho_{\rm{m,0}}$ is the component of the critical density of the universe in matter at $z=0$, and $\sigma\left(R,z\right)$ is the variance of the matter density field on the scale $R$. The variance depends on the linear matter power spectrum $P\left(k,z\right)$ through
	\begin{equation}
	\label{eqn:variance}
	\sigma^{2}\left(R,z\right) = \frac{1}{2 \pi^2}\int_{0}^{\infty} k^2 P\left(k,z\right) |\tilde{W}\left(kR\right)|^{2} dk
	\end{equation}
	where $\tilde{W}\left(kR\right)$ is the Fourier transform of the spherical top-hat window function. The parameter $c_0$ anchors the normalization to that of a $10^{8} \msun$ halo at $z=0$. We introduce the factor $\left(1+z\right)^{\zeta}$ to account for additional redshift evolution, similar to the empirical approach employed by \citet{Prada++12}. We marginalize over a Gaussian prior on $\zeta$ with mean -0.25 and variance 0.05, which tracks the redshift evolution of the theoretical mass-concentration relations. 
	
	In Figure \ref{fig:mcrelations} we show the mass-concentration relation in Equation \ref{eqn:mcrelation} as blue and green curves alongside several models from the literature \citep{Bullock++01,Prada++12,Ludlow++16,DiemerJoyce19}. 
	
	\subsection{Mass-concentration relation for subhalos and the subhalo mass function}
	
	As soon as a field halo is accreted into a more massive host, it becomes a subhalo and ceases to evolve through `pseudo-evolution', which refers to changing halo concentrations due to the evolving background density of the universe while the density normalization and scale radius remain fixed in physical coordinates \citep{Diemer++13}. For subhalos, the concentration defined in terms of $r_{200}$ becomes an ill-defined concept, and the structure of the subhalo evolves through tidal stripping effects that alter the density profile \citep[e.g.][]{Errani++17}. A complete prescription for subhalo density profiles requires a model for how tidal stripping evolves a subhalo whose physical parameters are determined at the time of infall \citep[e.g.][]{GreenvandenBosch19}. Without a detailed prescription for this effect implemented at the present time, and given the need for the number of free parameters to match the statistical constraining power of the current sample size of lenses, we simply evaluate subhalo concentrations at the time of infall when the mass-concentration relation in Equation \ref{eqn:mcrelation} is valid. To this end, we sample a probability density for the infall redshift as a function of halo mass and the main deflector redshift using the semi-analytic modeling code {\tt{galaticus}} \citep{Benson12}. 
	
	We render subhalos from a mass function parameterized as 
	\begin{equation}
	\label{eqn:subhalomfunc}
	\frac{d^2 N_{\rm{sub}}}{dm dA} =  \frac{\Sigma_{\rm{sub}}}{m_0} \left(\frac{m}{m_0}\right)^{\alpha} \mathcal{F} \left(M_{\rm{halo}}, z\right),
	\end{equation}
	where the scaling function $\mathcal{F} \left(M_{\rm{halo}}, z\right)$ 
	\begin{equation}
	\label{eqn:scaling}
	\log_{10} \left(\mathcal{F}\right) = k_1 \log_{10} \left(\frac{M_{\rm{halo}}}{10^{13} \msun}\right) + k_2 \log_{10}\left(z+0.5\right)
	\end{equation}
	accounts for evolution of the differential projected number density of subhalos with host halo mass $M_{\rm{halo}}$ and redshift. The fit $k_1 = 0.88$ and $k_2 = 1.7$ is determined from a suite of simulated host halos and their substructure generated with {\tt{galacticus}}. We add a tidal truncation radius to subhalo density profiles that depends on the mass of the subhalo and its position inside the host halo \citep{Gilman++19b}. 
	
	\subsection{Where does the lensing signal come from?}
	
	We show the magnification cross section for a $10^{8} \msun$ halo as a function of its concentration in Figure \ref{fig:cross}. The maximum magnification perturbation from a halo with $c=8$ is $10\%$, while the perturbation from a halo with $c = 22$ reaches $30 \%$. More concentrated halos will increase the frequency of flux-ratio perturbations relative to a population of low-concentration halos, as more concentrated halos are more efficient lenses. 
	
	\subsection{Forward modeling methodology}
	The most important conceptual feature of our Bayesian inference technique is the recognition that we may obtain posterior distributions of model parameters from simulated datasets generated with a forward model, circumventing the direct computation of an intractable likelihood function. The forward modeling technique detailed by \citet{Gilman++19b} simultaneously samples the dark matter quantities of interest and nuisance parameters such as the logarithmic slope of the main deflector mass profile $\gamma_{\rm{macro}}$ and the extent of the lensed background source $\sigma_{\rm{src}}$. Comparisons between the forward model output and the observed data are performed through the use of a summary statistic, which is used to estimate the likelihoods for each lens and compute the posterior. 
	
	We use a uniform prior on $c_0$ and $\beta$ between $1 - 30$ and $0.3 - 1.3$, respectively, a Gaussian prior on $\zeta$ with mean -0.25 and variance 0.05, and a Gaussian prior on the slope of the subhalo mass function $\alpha$ with mean (variance) -1.9 (0.025) \citep{Springel++08}. The priors on the additional parameters in Equations \ref{eqn:losmfunc} and \ref{eqn:subhalomfunc} are summarized in Table 2 of \citet{Gilman++19b}. 
	
	\section{Data}
	\label{sec:data}
	We use the image positions and flux ratios from eleven quadruply-imaged quasars to constrain the mass-concentration relation. Eight of these systems have flux ratios measured using narrow-line emission from the background quasar \citep{Nierenberg++14,Nierenberg++17,Nierenberg++19} and three with radio emission B0128+437, MG0414+0543, and PG 1115+080. The data for the radio systems are taken from \citet{Koopmans++03}, \citet{Katz++97}, and \citet{Chiba++05}, respectively. Both the narrow-line systems and radio lenses have background source sizes large enough ($\sim 1-60 \rm{pc}$) to avoid contaminating effects from micro-lensing, while retaining sensitivity to dark matter halos in the mass range $10^{6}-10^{9} \msun$. 
	
	We assume the population mean halo mass on $\log_{10} \left(M_{\rm{halo}}\right)$ of $13.3$ for B0128+437, and $13.0$ and $13.5$ for MG0414+0543 and PG 1115+080, respectively, each with variance 0.3 dex. We defer to \citet{Gilman++19b} for details. We model the luminous satellite galaxy visible near MG0414+0543 \citep{Ros++00} with a Gaussian prior on its Einstein radius $\mathcal{N} \left(0.2, 0.05\right)$ and also on its mass centroid, with astrometric uncertainties of 50 m.a.s. We use a uniform prior on the background source size between $1-25 \rm{pc}$ ($25-60 \rm{pc}$) for the radio (narrow-line) lenses.
	
	\section{Results}
	\label{sec:results}
	\begin{figure}
		\includegraphics[clip,trim=0cm 0cm 0cm
		0cm,width=0.45\textwidth,keepaspectratio]{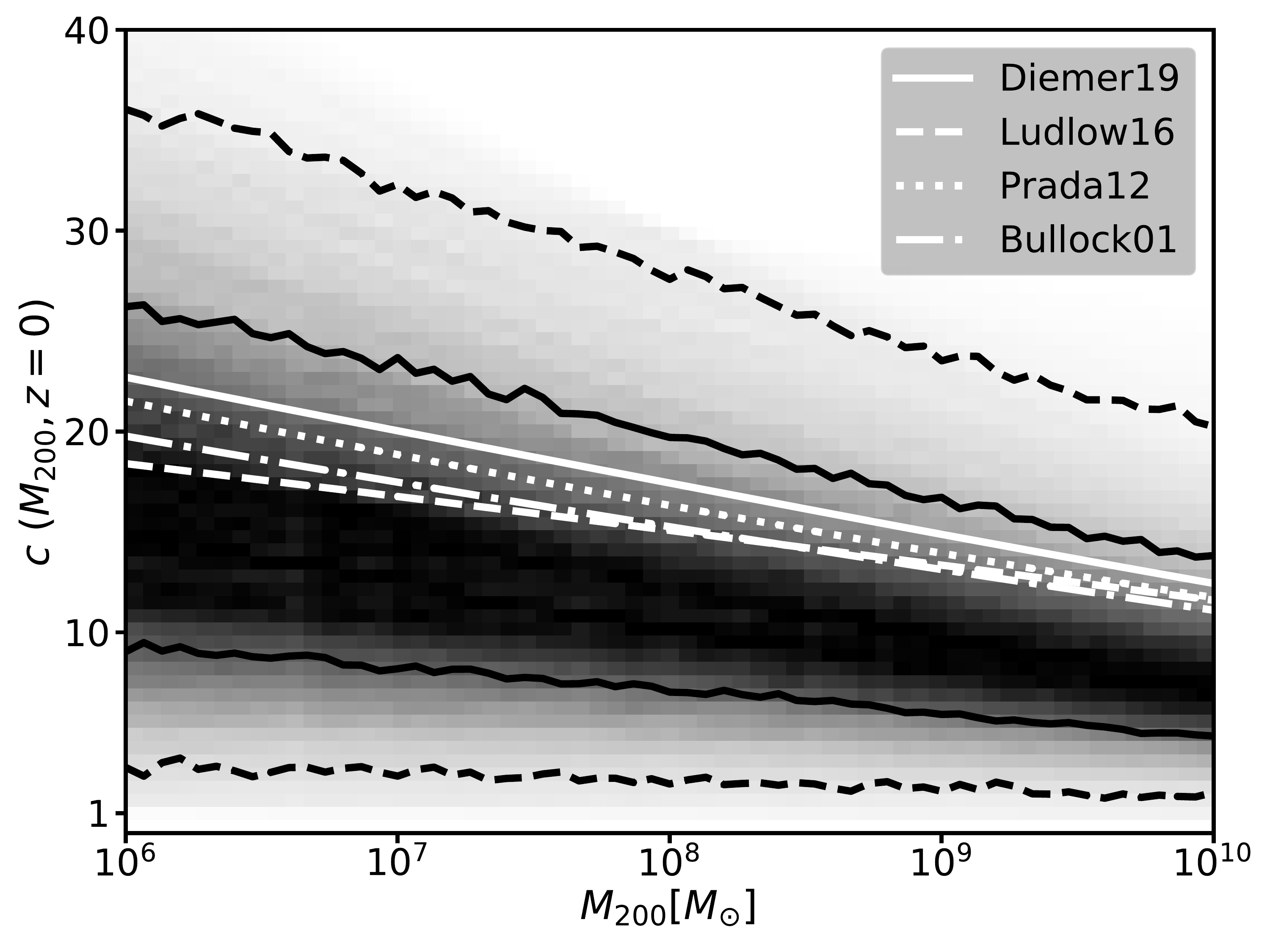}
		\caption{\label{fig:mcphys} Constraints on the concentration-mass relation of CDM halos derived from the posterior distribution of hyper-parameters shown in Figure \ref{fig:joint}, computed with eleven strong gravitational lenses. Black solid (dashed) lines contain $68\%$ ($95\%$) confidence intervals at fixed halo mass. White curves show several concentration-mass relations from the literature also plotted in Figure \ref{fig:mcrelations}.}
	\end{figure}	
	Our main results are the constraints on the hyper-parameters defining the mass-concentration relation in Equation \ref{eqn:mcrelation}. The posterior distributions for $c_0$, $\beta$, and $\Sigma_{\rm{sub}}$ are shown in Figure \ref{fig:joint}. The parameters $\beta$ and $\Sigma_{\rm{sub}}$ are relatively unconstrained due to the covariances present between $\Sigma_{\rm{sub}}$, $\beta$, and $c_0$. Both the normalization of the subhalo mass function $\Sigma_{\rm{sub}}$ and the normalization of the mass-concentration relation appear correlated with $\beta$. While the posterior distribution of $\beta$ is peaked towards higher values, the peak is not statistically significant. The parameters $c_0$ and $\Sigma_{\rm{sub}}$ are covariant, as more concentrated halos and more numerous halos both increase the clumpiness of dark matter structure on small scales. The parameter $\zeta$ is unconstrained. 
	
	The normalization of the mass-concentration relation is constrained by the data. The inference on the parameter $c_0$, which in our parameterizations is defined as the median concentration of a $10^{8} \msun$ halo at $z=0$ is at $1 \sigma$ \eightonesigma, and at $2 \sigma$ \eighttwosigma. This result is marginalized over the normalization of the subhalo mass function, the amplitude of the line of sight halo mass function, the slope of the subhalo mass function, and nuisance parameters describing the main deflector lens models and the background source size. 
	
	We translate the posterior distribution of hyper-parameters in Figure \ref{fig:joint} into constraints on the halo concentrations as a function of mass. The result is shown in Figure \ref{fig:mcphys}. Our constraints are consistent with the subset of models from the literature applicable in the halo mass range $10^6 - 10^{10} \msun$ relevant for our analysis. The uncertainties on concentrations of low-mass halos are larger than those of high mass halos, a result of the correlation between $c_0$ and $\beta$ in the posterior distribution in Figure \ref{fig:joint}. For a $10^{7} \msun$ halo, the $68 \%$ CI ($95 \%$ CI) constraints on the concentration parameter are \sevenonesigma (\seventwosigma). For a $10^{9} \msun$ halo, the $68 \%$ ($95 \%$) constraints on the concentration parameter are \nineonesigma (\ninetwosigma). 
	
	\section{Discussion and conclusions}
	\label{sec:conclusions}
	We have extended the Bayesian inference framework detailed by \citet{Gilman++19b} to accommodate a variable mass-concentration relation assuming a CDM mass function, and constrain the parameters describing this relation on sub-galactic scales using 11 quadruple-image strong gravitational lenses. Our main results are summarized as follows: 
	
	\begin{itemize}
		\item We constrain the normalization of the mass concentration relation $c_0$, defined as the concentration of a $10^{8} \msun$ halo at $z=0$. At $68 \%$ CI, \eightonesigma, and at $95 \%$ CI \eighttwosigma.
		\item We convert the constraints on the hyper-parameters describing the mass-concentration relation into physical halo concentrations as a function of halo mass. At $68 \%$ ($95 \%$), the concentration of a $10^{7} \msun$ halo is \sevenonesigma (\seventwosigma), while for a $10^{9} \msun$ halo \nineonesigma (\ninetwosigma). 
	\end{itemize}
	
	The results of this paper conclusively establish strong gravitational lensing by galaxies as perhaps the only probe of the mass-concentration relation of dark matter halos across cosmological distance on mass scales where they are expected to be completely or mostly dark. In future work, dedicated studies from N-body simulations and semi-analytic models will be necessary to refine theoretical predictions for the complicated processes that can shape the density profiles of dark matter subhalos in tidal fields. The sample size of strong lens suitable for the kind of study carried out in this work will increase by order of magnitude \citep{Oguri+10,Treu++18} in the coming decade, allowing strong lensing to constrain these processes. 
	
	\section*{Acknowledgments}
	We thank James Bullock and Benedikt Diemer for interesting discussions. 
	
	DG, TT, and SB acknowledge support by the US National Science Foundation through grant AST-1714953. DG, TT, SB and AN acknowledge support from HST-GO-15177. AJB and XD acknowledge support from NASA ATP grant 17-ATP17-0120. Support for Program number GO-15177 was provided by NASA through a grant from the Space Telescope Science Institute, which is operated by the Association of Universities for Research in Astronomy, Incorporated, under NASA contract NAS5-26555. TT and AN acknowledge support from HST-GO-13732. AN acknowledges support from the NASA Postdoctoral Program Fellowship, the UC Irvine Chancellor's Fellowship, and the Center for Cosmology and Astroparticle Physics Fellowship. 
	
	\bibliographystyle{mnras}
	\bibliography{bibliography}
	
\end{document}